\documentclass{PoS}
\title{
Calculation of $K \to \pi\pi$ decay amplitudes
with improved Wilson fermion
}
\ShortTitle{
Calculation of $K \to \pi\pi$ decay amplitudes
with improved Wilson fermion
}
\author{
\speaker{N.~Ishizuka}${}^{a,b}$\thanks{E-mail : ishizuka@ccs.tsukuba.ac.jp},
K.-I.~Ishikawa ${}^{c  }$,
A.~Ukawa       ${}^{a,b}$,
T.~Yoshi\'e    ${}^{a,b}$
\\ \\ \\
\llap{$^a$}
Graduate School of Pure and Applied Sciences,
University of Tsukuba, Tsukuba 305-8571, Japan
\\
\llap{$^b$}
Center for Computational Sciences,
University of Tsukuba, Tsukuba 305-8577, Japan
\\
\llap{$^c$}
Department of Physics, Hiroshima University,
Higashi-Hiroshima 739-8526, Japan
}
\FullConference{
31st International Symposium on Lattice Field Theory LATTICE 2013 \\
July 29 - August 3, 2013 \\
Mainz, Germany
}
\abstract{
We present results of our trial calculation
of the $K \to \pi\pi$ decay amplitudes
with the improved Wilson fermion action.
Calculations are carried out with
$N_f=2+1$ gauge configurations generated
with the Iwasaki gauge action and non-perturbatively
$O(a)$-improved Wilson fermion action
at $a=0.091\,{\rm fm}$, $m_\pi=280\,{\rm MeV}$ and $m_K=560\,{\rm MeV} (\sim 2 m_\pi)$
on a $32^3\times 64$ ($La=2.9 {\rm fm}$) lattice.
}
\begin{document}
%
%
\section{ Introduction }
Calculation of the $K\to\pi\pi$ decay amplitudes
is very important
to quantitatively understand the $\Delta I=1/2$ rule
in the neutral $K$ meson system and
the prediction of the direct $CP$ violation parameter ($\epsilon'/\epsilon$)
from the standard model.
A result for the decay amplitude for the $\Delta I=3/2$ process
at the physical quark mass was reported by RBC-UKQCD Collaboration
in Ref.~\cite{A2:RBC-UKQCD}.
They also reported a first direct calculation of the amplitude
for the $\Delta I=1/2$ process
carried out at $m_\pi=422\,{\rm MeV}$ in Ref.~\cite{A0:RBC-UKQCD}.
They employed the domain wall fermion action in these calculations.

In the present work
we attempt a direct calculation
of the $K\to\pi\pi$ decay amplitudes
for both the $\Delta I=1/2$ and $3/2$ processes
with the improved Wilson fermion action.
As we discuss below, mixings with four-fermion operators
with  wrong chirality are absent
for the parity odd process even for the Wilson fermion action.
A mixing to a lower dimension operator does occur,
which gives unphysical contributions to the amplitudes on the lattice.
However,  it can be non-perturbatively subtracted
by imposing a renormalization condition.
After the subtraction we can obtain the physical decay amplitudes
by the renormalization factor
having the same structure as for the continuum.
Therefore, by using the Wilson fermion action,
statistical improvement is expected with the lattice calculation
of the amplitudes for the $\Delta I=1/2$ process,
because calculations with Wilson fermion action
are computationally much less expensive
than those with the domain wall fermion action.

Our calculations are carried out on a subset of
configurations previously generated by PACS-CS Collaboration
with the Iwasaki gauge action and non-perturbatively
$O(a)$-improved Wilson fermion action at $\beta=1.9$
on a $32^3\times 64$ lattice~\cite{conf:PACS-CS}.
The subset corresponds to the hopping parameters
$\kappa_{ud}=0.13770$ for the up and down quark and
$\kappa_{s }=0.13640$ for the strange quark.
The parameters determined from the spectrum analysis
for this subset are $a=0.091\ {\rm fm}$ and $La=2.91\ {\rm fm}$,
$m_\pi=280\,{\rm MeV}$ and $m_K=560\,{\rm MeV} (\sim 2 m_\pi)$.
We consider the $K$ meson decay process to the zero momentum two pions
on these configurations.
We further generate gauge configurations
at the same parameters to improve the statistics.
The total number of  gauge configurations used in the present work is $343$
which corresponds to $8,575$ trajectories.
%
%
\section{ Operator mixing }
In this section
we discuss operator mixing of the $\Delta S=1$ weak operators
for the Wilson fermion action.
In the continuum,
the effective Hamiltonian of the $K\to\pi\pi$ decay is given by
a linear combination of 10 four-fermion operators
($Q_j$ for $j=1,2,\cdots 10$)~\cite{Review:Buras}, of which
7 operator are linearly independent.
They can be classified
by the irreducible representation
of the flavor ${\rm SU}(3)_L \times {\rm SU}(3)_R$ symmetry group
as $({\bf 27},{\bf 1}) + 4\cdot ({\bf 8},{\bf 1}) + 2\cdot ({\bf 8},{\bf 8})$,
whose components are given by
\begin{equation}
\begin{array}{lll}
({\bf 27},{\bf 1})
&   Q_1' = 3 Q_1 + 2 Q_2 - Q_3 \ , \\
({\bf 8},{\bf 1})
&   Q_2' =   2 Q_1 - 2 Q_2 + Q_3 \ ,
&   Q_3' = - 3 Q_1 + 3 Q_2 + Q_3 \ , \\
&  Q_5 = (\bar{s}       d) (\bar{u}       u + \bar{d}       d + \bar{s}       s )_{LR} \ ,
&  Q_6 = (\bar{s}\times d) (\bar{u}\times u + \bar{d}\times d + \bar{s}\times s )_{LR} \ ,  \\
({\bf 8},{\bf 8})
&  Q_7 = (\bar{s}       d) (\bar{u}       u - \bar{d}       d/2 - \bar{s}       s/2 )_{LR} \ , \quad
&  Q_8 = (\bar{s}\times d) (\bar{u}\times u - \bar{d}\times d/2 - \bar{s}\times s/2 )_{LR} \ , \\
\end{array}
\label{eq:op}
\end{equation}
with $Q_1=(\bar{s}d)(\bar{u}u)_{LL}$, $Q_2=(\bar{s}\times d)(\bar{u}\times u)_{LL}$
and  $Q_3=(\bar{s}d)(\bar{u}u+\bar{d}d+\bar{s}s)_{LL}$,
where $(\bar{s}d)(\bar{u}u)_{L,R/L}
= ( \bar{s} \gamma_\mu ( 1 - \gamma_5 ) d )( \bar{u} \gamma_\mu ( 1 \pm \gamma_5 ) u)$
and $\times$ means contraction of the color indices :
$
    (\bar{s}\times d)_L (\bar{u}\times d)_L
  = (\bar{s}_a   d_b)_L (\bar{u}_b d_a)_L
$.

In the continuum, mixings between operators in different representations are forbidden.
For the Wilson fermion action, however,
chiral symmetry is broken to the vector subgroup,
${\rm SU}(3)_L \times {\rm SU}(3)_R \to {\rm SU}(3)_V$.
Hence mixings among different representations are in general allowed,
and new operators arise through radiative corrections.
However, these problems are absent for the parity odd part
of the operators in (\ref{eq:op}),
which are the operators considered
for the direct calculation of the $K\to\pi\pi$ decay amplitudes
in the present work.

To investigate the operator mixing,
we exploit the full set of unbroken symmetries for the Wilson fermion, namely
flavor ${\rm SU}(3)_V$, parity $P$, charge conjugation $C$,
and $CPS$ which is the symmetry under $CP$ transformation
followed by the exchange of the $d$ and $s$ quarks.
All operators in (\ref{eq:op}) are  $CPS=+1$ operators.
We know that the following operators also have the same quantum numbers
as the operators in (\ref{eq:op}).  We therefore
have to consider operator mixing with them,
\begin{equation}
 Q_X = (\bar{s}       d) (\bar{d}       d - \bar{s}       s )_{SP+PS}  \quad , \quad
 Q_Y = (\bar{s}\times d) (\bar{d}\times d - \bar{s}\times s )_{SP+PS}
\ ,
\label{eq:op_2}
\end{equation}
where
$ (\bar{s}d)(\bar{d}d )_{SP+PS}
= (\bar{s}d)_S (\bar{d}d )_P + (\bar{s}d)_P (\bar{d}d )_S$,
$(\bar{s}d)_S = \bar{s} d$ and
$(\bar{s}d)_P = \bar{s} \gamma_5 d$.

It was shown in Ref.~\cite{CPS:Donini} that
the parity odd part of the $LL$ and $LR$ type operators,
and the $SP+PS$ type operators do not mix
with each other by the gluon exchange diagrams
due to the $CPS$, $CPS'$ and $CPS''$ symmetry,
where $S'$ is defined as
$(\overline{\psi_1}\psi_2)(\overline{\psi_3}\psi_4)\to
 (\overline{\psi_2}\psi_1)(\overline{\psi_4}\psi_3)$
and $S''$ by
$(\overline{\psi_1}\psi_2)(\overline{\psi_3}\psi_4)\to
 (\overline{\psi_4}\psi_3)(\overline{\psi_2}\psi_1)$.
Thus the operators $Q_{X,Y}$ (the $SP+PS$ type)
do not mix with those in (\ref{eq:op}) (the $LL$ and $LR$ type).
The operators $Q_{7,8}\in ({\bf 8},{\bf 8})$ (the $LR$ type)
do not mix with the $LL$ type operators
($Q_{1,2,3}\in({\bf 27},{\bf 1}), ({\bf 8},{\bf 1})$),
and also with $Q_{5,6}\in({\bf 8},{\bf 1})$ (the $LR$ type)
because the gluon exchange diagrams do not change the flavor structure
and these operators have different structures.
Further the mixing between the $({\bf 27},{\bf 1})$
and the $({\bf 8},{\bf 1})$ representation
is forbidden by the flavor ${\rm SU}(3)_V$ symmetry.
To sum up, the renormalization factor for the gluon exchanging diagrams
has the same form as for the continuum.

Next let us investigate the possibility of unwanted mixings
though the penguin diagrams.
In the penguin diagrams for $Q_{7,8} \in ({\bf 8},{\bf 8})$,
cancellation of the quark loops at the weak operators occurs as $d-s$.
This means that the renormalization due to the penguin diagram
is proportional to the quark mass difference and mixing to four-fermion operators
is absent due to the dimensional reason.
In addition the operator arising from the penguin diagrams should have
the  flavor structure
$(\bar{s}d)(\bar{u}u + \bar{d}d + \bar{s}s )$,
which is different from that of $Q_{7,8}$.
Thus operator mixing from $Q_{7,8} \in ({\bf 8},{\bf 8})$ to the other representations
and its reverse are absent.
These statements also hold for $Q_{X,Y}$ in (\ref{eq:op_2}) for the same reason,  and
the operators $Q_{X,Y}$ are fully isolated in the theory.

Up to now, we have shown that
the renormalization factor
for the parity odd part of the four-fermion operators in (\ref{eq:op})
have the same form as that in the continuum.
Here we consider the mixing to  lower dimensional operators.
From the $CPS$ symmetry and the equation of motion of the quark,
there is only one operator with ${\rm dim}<6$, which is
\begin{equation}
   Q_P = ( m_d - m_s ) \cdot P = ( m_d - m_s ) \cdot \bar{s} \gamma_5 d
\ .
\label{eq:Q_P}
\end{equation}
This operator also appears in the continuum,
but does not give a finite contribution to the physical decay amplitude,
since it is a total derivative operator.
But this is not valid for the Wilson fermion due to  chiral symmetry breaking by the Wilson term,
and the operator (\ref{eq:Q_P}) gives a non-zero unphysical contribution
to the amplitudes on the lattice.
This contribution should be subtracted non-perturbatively,
because the mixing coefficient includes
a power divergence due to the lattice cutoff growing as  $1/a^2$.
In the present work
we subtract it by imposing the following relation~\cite{sub:Dawson},
\begin{equation}
    \langle 0 | \, \overline{Q}         \,| K \rangle
 =  \langle 0 | \, Q - \alpha(Q)\cdot P \,| K \rangle = 0
\ ,
\label{eq:sub_Q}
\end{equation}
for  each operator in (\ref{eq:op}).
The subtracted operators $\overline{Q}$ are renormalized
by the renormalization factor having same form as in the continuum.
%
%
\section{ Calculation }
We extract the decay amplitude from the time correlation function of
the $K\to\pi\pi$ process,
\begin{equation}
  G(Q^I)(t) = \frac{1}{T} \sum_{\delta=0}^{T-1}
             \langle 0 | \, W_K(t_K+\delta) \,\, \overline{Q}(t+\delta) \,\,
                            W_{\pi\pi}^{I}(t_\pi+\delta) \, | 0 \rangle
\ ,
\label{eq:G_KPIPI}
\end{equation}
where $W_K(t)$ is the wall source for the $K^0$ meson
and $W_{\pi\pi}^{I}(t)$ is that for the isospin $I$ two-pion system.
$\overline{Q}(t)$ is the subtracted weak operator defined by (\ref{eq:sub_Q}).
We impose the periodic boundary condition in all directions.
The summation over $\delta$, 
where $T=64$ denotes the temporal size of the lattice,
is taken to improve the statistics.
We set $t_K=26$ and $t_\pi=0$ in the present work.
The gauge configurations are fixed to the Coulomb gauge 
at the time slice of the wall source $t=t_K+\delta$ and $t_\pi+\delta$
for each $\delta$.
The mixing coefficient of the lower dimensional operator
$\alpha(Q)$ is evaluated from the ratio,
\begin{equation}
\alpha(Q) =
\sum_{\delta_1=0}^{T-1} \langle 0 | \, W_K(t_K+\delta_1) \, Q(t+\delta_1) \, | 0 \rangle
\Bigl/ \,
\sum_{\delta_2=0}^{T-1} \langle 0 | \, W_K(t_K+\delta_2) \, P(t+\delta_2) \, | 0 \rangle
\ ,
\end{equation}
in  the large $t$ region.

For the calculation of quark loop at the weak operator $Q(x,x)$, {\it i.e.},
the quark propagator starting from the weak operator
and ending at the same position,
we use the stochastic method with
the hopping parameter expansion technique (HPE)
and the truncated solver method (TSM)
proposed in Ref.~\cite{HPE_RSM:Bali}.
The action of the Wilson fermion can be written as
\begin{eqnarray}
&&
  S^{W}
    = \bar{\psi}\, W        \, \psi
    = \bar{\psi}\, ( M - D )\, \psi
    = \bar{\psi}\, M ( 1 - \bar{D} )\, \psi
           \quad , \qquad( \,\, \bar{D}= M^{-1} D \,\, )
\label{eq:action_W}
\\
&&
\quad
   ( M \psi )(x) = ( 1 - \kappa C_{SW} (\sigma \cdot F(x) )/2 ) \psi(x)
\ ,   \\
&&
\quad
   ( D \psi )(x)
   = \kappa \sum_\mu
        \left(   p^{-}_{\mu}U_\mu(x) \psi(x+\mu)
               + p^{+}_{\mu}U_\mu^\dagger(x-\mu) \psi(x-\mu) \right)
\ ,
\end{eqnarray}
where $p_\mu^{\pm} = 1 \pm \gamma_\mu$.
From (\ref{eq:action_W}) the quark propagator $Q$ can be written
by a hopping parameter expansion form as
$Q = W^{-1}
    = \sum_{n=0}^{k-1} \bar{D}^n M^{-1} + \bar{D}^k W^{-1}
$
for any integer value of $k$.
We use this to calculate the quark loop $Q(x,x)$ at the weak operator.
In this case, the terms with the odd power of $\bar{D}$
do not contribute,
thus $Q(x,x)=( M^{-1} + \bar{D}^2 M^{-1} + \bar{D}^4 W^{-1} )(x,x)$ for $k=4$.
Using this,
we calculate the quark loop by the stochastic method as,
\begin{eqnarray}
&&
  Q({\bf x},t ; {\bf x},t ) =
    \frac{1}{N_R}
    \sum_{j=1}^{N_R}\,
         \xi^{*}_j({\bf x},t) \,\, S_j({\bf x},t)
\ ,
\label{eq:HPE_stoch}
\\
&&
\qquad
  S_j({\bf x},t)
   =  \sum_{\bf y} \,
      \bigl( M^{-1} + \bar{D}^2 M^{-1} + \bar{D}^4 W^{-1} \bigr)({\bf x}, t; {\bf y}, t )
      \,\,  \xi_j({\bf y},t)
\ ,
\label{eq:HPE_stoch_S}
\end{eqnarray}
where
we introduce $U(1)$ noise $\xi_j({\bf x},t)$ which satisfies
$
  \delta^3 ( {\bf x} - {\bf y} ) =
  1/N_R \cdot
  \sum_{j=1}^{N_R}
       \xi_j^{*} ({\bf x},t) \xi_j ({\bf y,t})
$
for $N_R \to \infty$.
The effect of HPE for the quark loop
is removing the $\bar{D}$ and $\bar{D}^3$ terms in (\ref{eq:HPE_stoch_S}) explicitly
which make only statistical noise.
We find that
HPE reduce the statistical error of the decay amplitudes
to about $50\%$ compared with the normal stochastic method.

We also implement the truncated solver method (TSM)
for (\ref{eq:HPE_stoch}) by
\begin{equation}
  Q({\bf x},t ; {\bf x},t ) =
  \frac{1}{N_T}
  \sum_{j=1}^{N_T}\,
       \xi^{*}_j({\bf x},t) \,\, S_j^T({\bf x},t)
  \,\,
  +
  \,\,
  \frac{1}{N_R}
  \sum_{j=N_T+1}^{N_T+N_R}\,
       \xi^{*}_j({\bf x},t) \,\, \bigl[ S_j({\bf x},t) - S_j^T({\bf x},t) \bigr]
\ ,
\label{eq:TSM}
\end{equation}
where
$S^T_j({\bf x},t)$ is a value given with the quark propagator $W^{-1}$
calculated with a loose stopping condition in (\ref{eq:HPE_stoch_S})
and $S_j({\bf x},t)$ is that with a stringent condition.
We set $N_T=5$ and the stopping condition
$R\equiv | W W^{-1} - \xi |/|\xi| < 1.2 \times 10^{-6}$ for $S_j^T({\bf x},t)$,
and $N_R=1$ and $R < 10^{-14}$ for $S_j({\bf x},t)$.
We find that
contributions of the second term of (\ref{eq:TSM})
to the decay amplitudes
are negligible compared with the statistical error.
Thus we neglect the second term in (\ref{eq:TSM})
and estimate the quark loop by only the first term
by setting $N_T=6$ for TSM, 
confierming that 
the contributions of the second term are negligible
by additional calculations of $S_j({\bf x},t)$
for all gauge configurations.
The numerical cost of TSM (\ref{eq:TSM})
is about twice of that without TSM (\ref{eq:HPE_stoch}) with $N_R=1$.
%
%
\section{ Results }
%
%
\begin{figure}[t]
\begin{center}
\includegraphics[width=87mm]{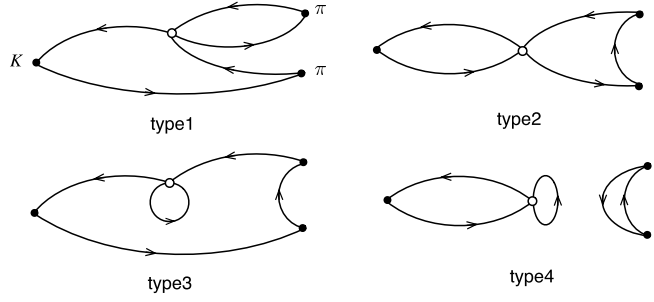}
\end{center}
\vspace{-6mm}
\caption{
Quark contraction of $K\to\pi\pi$ decay.
}
\label{fig:cont}
\end{figure}
\begin{figure}[p]
\begin{center}
\includegraphics[width=130mm]{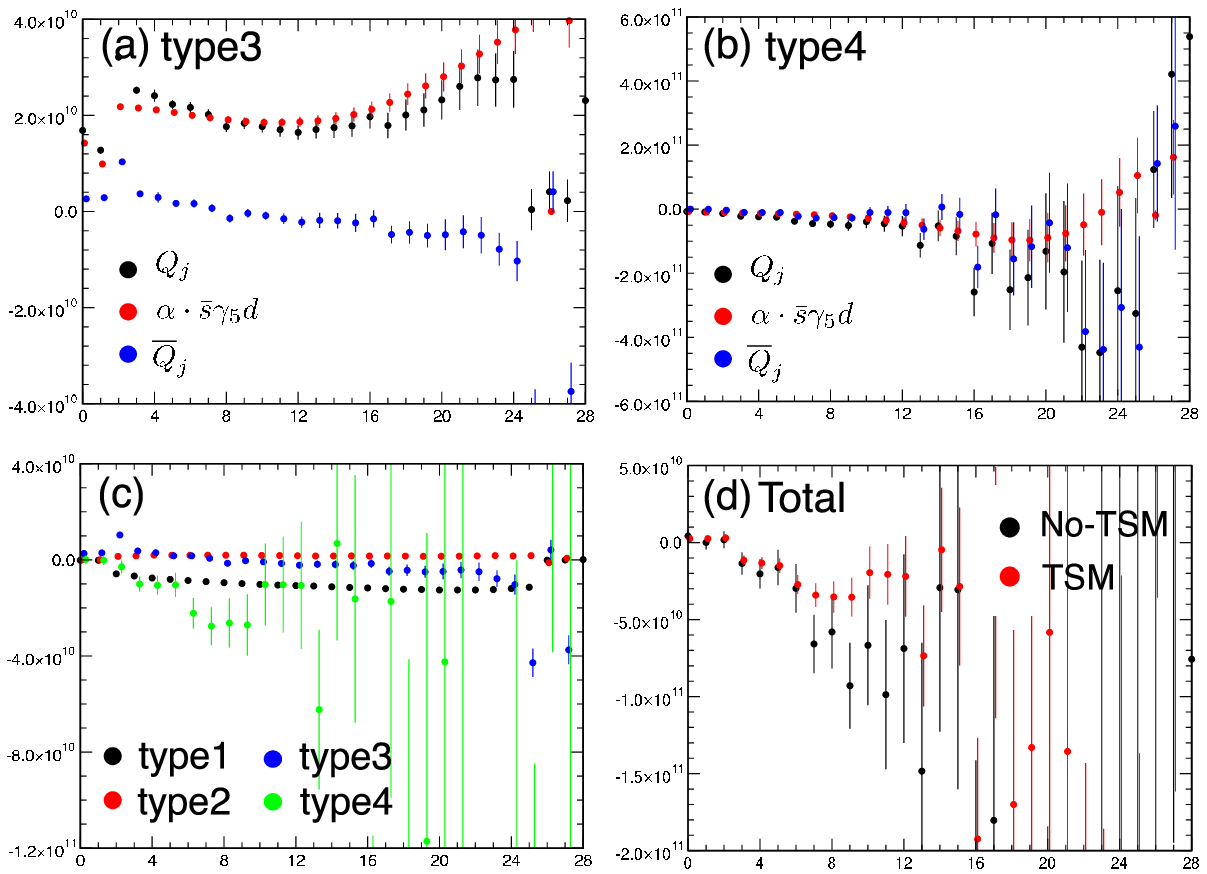}
\end{center}
\vspace{-3mm}
\caption{
Time correlation function
of $Q_2$ for the $\Delta I=1/2$ decay.
(a) {\it type3} contribution for $Q_2$, $\alpha(Q_2)\cdot P$ and
$\overline{Q}_2 = Q_2 - \alpha(Q_2)\cdot P$,
(b)  {\it type4} contribution,
(c) results for each type of contractions for $\overline{Q}_2$,
(d) results of the total correlation functions calculated with TSM and without TSM.
}
\label{fig:Q2I0}
%
%
\begin{center}
\includegraphics[width=130mm]{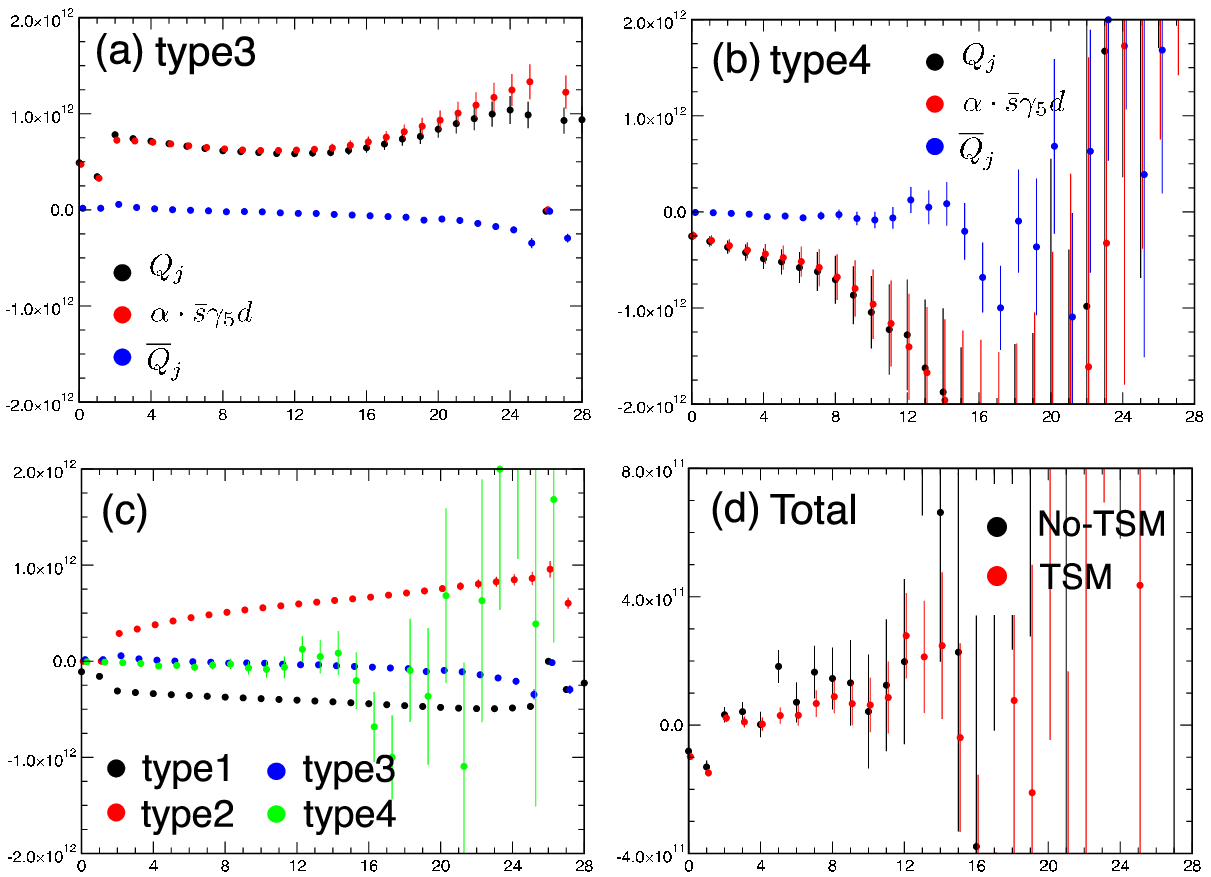}
\end{center}
\vspace{-3mm}
\caption{
Time correlation function
of $Q_6$ for the $\Delta I=1/2$ decay
following the same convention as in Fig.~\protect\ref{fig:Q2I0}.
}
\label{fig:Q6I0}
\end{figure}

%
%
There are four types of quark contractions
for the $K\to\pi\pi$ decay as shown in Fig.~\ref{fig:cont},
where the naming of the contractions follows that 
by RBC-UKQCD~\cite{A0:RBC-UKQCD}.
The results for the time correlation function
(\ref{eq:G_KPIPI}) of $Q_2$ for the $\Delta I=1/2$ process
are plotted in Fig.~\ref{fig:Q2I0}.
We adopt $K^0 = - \bar{d} \gamma_5 s$ as the neutral $K$ meson operator,
so our correlation function has an extra minus from the usual convention.
We find a large cancellation in $\overline{Q}_2$
between the contributions from the operator $Q_2$ and
$\alpha(Q_2)\cdot P$ for the {\it type3} contraction.
This is not seen for the {\it type4} contraction.
In (d) we compare the correlation functions
calculated with TSM (\ref{eq:TSM})
and without TSM (\ref{eq:HPE_stoch}) with $N_R=1$.
TSM improves the statistics drastically.
The numerical cost of TSM is about twice of
that without TSM as already mentioned.
Thus TSM is a very efficient method.
%
%
%
%

The results for  $Q_6$ for $\Delta I=1/2$
are shown in Fig.~\ref{fig:Q6I0}.
Here we find a large cancellation in $\overline{Q}_6$ between
the contributions of $Q_6$ and $\alpha(Q_6)\cdot P$
for both the {\it type3} and {\it type4} contractions.
In (c) a large cancellation is also seen between
the {\it type1} and  {\it type2} contractions.
An efficiency of TSM is also observed for $Q_6$ in (d).

We extract the decay amplitude
$M(Q^I)= \langle K |\, \overline{Q}\, | \pi\pi; I \rangle$
by fitting the time correlation function (\ref{eq:G_KPIPI})
with a fitting function,
\begin{equation}
  G(Q^I)(t) = M(Q^I) \cdot
   ( 1 / F_{LL} )\cdot N_K N_{\pi\pi}^{I}
   \cdot {\rm e}^{ - m_K (t_K-t) - E_{\pi\pi} (t-t_\pi) }
   \times (-1)
\ ,
\label{eq:amp_M}
\end{equation}
with the energy of the two-pion state $E_{\pi\pi}$
which is fixed at a value
obtained from the $\pi\pi\to\pi\pi$ correlation function.
The factor $(-1)$ comes from the convention of the $K^0$ operator.
The factor
$N_K = \langle 0 | W_K | K \rangle$ and
$N_{\pi\pi}^{I} = \langle 0 | W_{\pi\pi}^{I} | \pi\pi; I \rangle$
are estimated from the wall to wall propagator of the $K$ meson and the two-pion.
$F_{LL}$ is the Lellouch-L\"uscher factor~\cite{LL-factor:LL}.
In the present work,
precise results for the scattering phase shift
for the two-pion system are not yet available,
so we adopt the factor for the noninteracting case,
$F_{LL}=\sqrt{ 2 m_K L^3} ( \sqrt{ 2 m_\pi L^3 } )^2$.
Our results are given by
\begin{equation}
a \cdot M(Q_2^{I=0}) = ( + 4.43 \pm 1.62 ) \times 10^{-2} \ , \quad
a \cdot M(Q_6^{I=0}) = ( - 1.34 \pm 0.85 ) \times 10^{-1} \ ,
\end{equation}
for the fitting range $t=[8,12]$.
The signal to noise ratios are comparable to those of RBC-UKQCD~\cite{A0:RBC-UKQCD}.
In the next step,
we will correct these bare values by the renormalization factors and multiply with the
coefficient functions to obtain the physical results for $A_0$ and $A_2$.
%
%
\section{ Summary }
In the present work
we have reported on our attempt at a direct calculation
of the $K\to\pi\pi$ decay amplitudes for both the $\Delta I=1/2$ and $3/2$ channels
with the Wilson fermion action.
We have shown that the unwanted mixings with wrong chirality operators
are absent for the $K\to\pi\pi$ decay even for the Wilson fermion action.
We have calculated the decay amplitudes
by using the stochastic method with the hopping parameter expansion technique
and the truncated solver method.
We have shown that these two methods are efficient.
%
%
\section*{Acknowledgments}
This work is supported in part by Grants-in-Aid
of the Ministry of Education No.~23340054.
The numerical calculations have been carried out
on T2K-TSUKUBA at University of Tsukuba,
SR16000 at University of Tokyo,
and K-computer at RIKEN AICS.
%
%

%
%
\end{document}